\documentstyle[12pt]{article}
\begin{document}

\par
\vskip .5 truecm
\large \centerline{\Large{\bf Quantum Parafermions in the
$SL(2,R)/U(1)$}}
 \centerline{\Large{\bf WZNW Black Hole Model}}
 
\par
\vskip 1 truecm
\normalsize
\begin{center}
{\bf C.~Ford}\\
\sl{DESY Zeuthen, Platanenallee 6, Zeuthen 15738, Germany\\ 
\tt ford@ifh.de\rm\\
}
\end{center}

\begin{abstract}
Starting with its classical parafermion algebra, we
 consider the quantisation of the $SL(2,R)/U(1)$ WZNW black hole model.

\end{abstract}

\section{Introduction}

About ten years ago it was realised that certain $2d$ conformal field
theories 
have a black-hole interpretation.
 Witten \cite{witten}
pointed out that a non-nilpotently
gauged $SL(2,R)$ Wess Zumino Novikov Witten (WZNW)
theory describes a two dimensional Euclidean black hole.
He also suggested that the theory \it may be integrable \rm
opening up the possibility of an exact quantisation of the black hole
model.
Gervais and Saveliev \cite{gs}
 went one step further.
They noted that certain non-Abelian Toda
theories also represent black holes.
The importance of this observation is that the integrability of such
(\it nilpotently-gauged \rm WZNW) theories was not in doubt.
Furthermore, even though these are $2d$ conformal field theories
they can correspond to \it higher \rm dimensional black holes.
More precisely, certain nilpotently gauged $B_n$ WZNW theories 
contain
 $n$-dimensional black holes.
With a view to performing an exact canonical quantisation
Bilal \cite{bilal}
considered the simplest case, $B_2\equiv SO(5)$,
 in some detail.
The theory can be described via the Lagrangian
\begin{equation}\label{nonablag}
\gamma^2{\cal L}=
\partial_z r\partial_{\bar z}r+
\tanh^2 r \partial_z t\partial_{\bar z}t+
\partial_z\phi\partial_{\bar z}\phi
+\cosh(2r)e^{2\phi},
\end{equation}
where $r$ is a positive scalar field,
$t$ an `angular' field in that
$t$ and $t+2\pi$ are identified
and $\phi$ is a real scalar field.
$z=\tau+\sigma$ and $\bar z=\tau-\sigma$
are light cone coordinates
and
$\gamma$ is a coupling constant.
Using a general construction of Gervais and Savaliev
one can  write down  explicitly the general solution
to the classical equations of motion.
 This can be recast as a canonical transformation (CT)
mapping the `physical' $r$, $t$, $\phi$ fields to three free
fields $\phi_1$, $\phi_2$
, $\phi_3$.
The next step would be to
`promote' the CT to a quantum-mechanical
operator identity exchanging physical and free fields.
Such calculations are well known from Liouville theory
\cite{BCT,OW}. 
However the CT considered in \cite{bilal} is \it much \rm
more complicated than its Liouville counterpart
(it takes several pages just to write it down).

The Lagrangian for the $SL(2,R)/U(1)$ model is simpler
\begin{equation}\label{blacklag}
\gamma^2{\cal L}=\partial_z r\partial_{\bar z}r+
\tanh^2 r \partial_z t \partial_{\bar z}t.
\end{equation}
In fact it is just the same  as the $B_2$  theory
without the `Liouville'  $\phi$
field.
One can associate to the Lagrangian (\ref{blacklag})
 a `target space' metric
\begin{equation}\label{tsmetric}
ds^2=(dr)^2+\tanh^2 r (d t)^2.
\end{equation}
It is this that Witten interpreted as a $2d$ Euclidean black hole.
One can regard the $B_2$ theory as  a black hole
`interacting' with Liouville `matter'.
Guided by the  solution of the  non-Abelian Toda theory
M\"uller and Weigt \cite{MW}
deduced the solution
 of the $SL(2,R)/U(1)$ model
and gave a CT mapping the physical $r$ and $t$
fields onto two free fields $\phi_1$ and $\phi_2$.
This work established that the theory is indeed integrable.
To express the solution it is convenient to introduce
the `Kruskal' coordinates
\begin{equation}\label{kruskal}
u=e^{it}\sinh r,\qquad
\bar u= e^{-it}\sinh r.
\end{equation}
The CT is
\begin{equation}\label{ct}
u=e^{i\gamma(\phi_2+\bar \phi_2)}\left[
e^{\gamma(\phi_1+\bar\phi_1)}(1+\Phi\bar\Phi)-\frac{1}{4}
e^{-\gamma(\phi_1+\bar\phi_1)}
+
\frac{i}{2}\left(\Phi e^{\gamma(\phi_1-\bar\phi_1)}
+\bar\Phi e^{-\gamma(\phi_1-\bar \phi_1)}\right)\right],
\end{equation}
where the $\phi_i\equiv \phi_i(z)$ and
$\bar\phi_i\equiv \phi_i(\bar z)$
denote respectively the chiral and anti-chiral
part of a free field.
The chiral
object $\Phi\equiv\Phi(z)$ is defined via the differential equation
\begin{equation}
\partial_z\Phi(z)=e^{-2\gamma\phi_1(z)}\partial_z \phi_2(z),
\end{equation}
and similarly for $\bar\Phi(\bar z)$.

\section{Parafermion algebra}

Underlying the integrability of $SL(2,R)/U(1)$
model is a set of `parafermionic' chiral fields
\begin{equation}
V_\pm =\frac{1}{\gamma^2}e^{\pm i\nu}
\left(\partial_z r\pm i\tanh r \partial_z t\right).
\end{equation}
$\nu$ is defined through the differential equations
\begin{equation}
\partial_z\nu=(1+\tanh^2 r)\partial_z t,\qquad
\partial_{\bar z}\nu=\cosh^{-2}r\partial_{\bar z}t.
\end{equation}
The integrability condition for these equations
is \it one \rm
of the equations of motion.
Using the other equation of motion it follows that the
$V_\pm$ are \it chiral, \rm  i.e. $\partial_{\bar z}V_\pm=0$.
Similarly, one can define a pair of anti-chiral parafermions.
The chiral component of the energy momentum tensor
has the  Sugawara form
\begin{equation}\label{sugawara}
T(z)=\gamma^2V_+(z)V_-(z).
\end{equation}
It turns out that
the $V_\pm(z)$ are much simpler when written
in terms of the free fields
\begin{equation}\label{freefieldparas}
V_\pm(z)=\frac{1}{\gamma}\left(\partial_z\phi_1(z)\pm i\partial_z\phi_2
(z)\right)
\exp\left(\pm2i\gamma\phi_2(z)\right).
\end{equation}
The parafermions satisfy a closed Poisson bracket (PB)
algebra.
For concreteness let us fix the boundary conditions
and the basic PB's of the free fields.
We take spacetime to be $S^1\times R$ so that we have periodicity
in the spatial direction
\begin{equation}
u(\sigma+2\pi,\tau)=u(\sigma,\tau).
\end{equation}
For the free fields we have the mode expansions
\begin{equation}\label{freemode}
\phi_k(z)=\frac{1}{2}q_k+\frac{z}{4\pi}p_k
+\frac{i}{\sqrt{4\pi}}\sum_{n\neq 0}\frac{a_n^{(k)}}{n}
e^{-inz},~~~
\bar\phi_k(\bar z)=
\frac{1}{2}q_k+\frac{\bar z}{4\pi}p_k+\frac{i}{\sqrt{4\pi}}
\sum_{n\neq 0}\frac{\bar a_n^{(k)}}{n}e^{-in\bar z}.
\end{equation}
These fields are \it not \rm periodic
\begin{equation}
\phi_k(z+2\pi)=\phi_k(z)+\frac{p_k}{2},~~~~~
\bar\phi_k(\bar z-2\pi)=\bar\phi_k(\bar z)-\frac{p_k}{2},
\end{equation}
where the $p_k$ are momentum zero modes.
However, the \it full \rm free fields,
$\phi_i(z)+\bar\phi_i(\bar z)$, are.
The Fourier coefficients satisfy the PB
relations
\begin{equation}\label{modealgebra}
\{q_k,p_l\}=\delta_{kl},~~~~
\{a_m^{(k)},a_n^{(l)}\}=-im\delta_{kl}\delta_{m+n\, 0},
~~~~\{q_k,a_m^{(l)}\}=\{p_k,a_m^{(l)}\}=0,
\end{equation}
and similarly for the $\bar a_n^{(k)}$'s.

Note that, like the $\phi_i$'s,
the $V_\pm$'s are not periodic.
When computing PB's and their quantum counterparts
we prefer to deal with periodic objects.
Thus instead of the $V_\pm$'s we will consider
\begin{equation}\label{periodicparas}
W_\pm(z)=\frac{1}{\gamma}\left(
\partial_z\phi_1(z)\pm i\partial_z\phi_2(z)\right)e^{
2i\gamma\varphi_2(z)},
\end{equation}
where $\varphi_2$ is $\phi_2$ with the momentum zero mode removed
and the whole zero mode of the full free field included,
i.e. $\varphi_2(z)=\frac{1}{2}q_2+
\phi_2(z)|_{p_2=0}$.
These periodic parafermions obey the non-linear
PB algebra
\begin{eqnarray}\label{PB1}
\{W_\pm(z),W_\pm(z')\}&=&\gamma^2
W_\pm(z)W_\pm(z')h(z-z'),\\ \label{PB2}
\{W_\pm(z),W_\mp(z')\}&=&\!\!\!-\gamma^2
W_\pm(z) W_\mp(z')h(z-z')+\frac{1}{\gamma^2}\left(
\partial_z+\frac{i\gamma p_2}{2\pi}\right)
\delta_{2\pi}(z-z'),\\ \label{PB3}
\{p_2,W_\pm(z')\}&=&\mp2i\gamma W_\pm(z'),
\end{eqnarray}
where
\begin{equation}
h(z)=\left(\epsilon_{2\pi}(z)-\frac{z}{\pi}\right),
\end{equation}
is the periodic sawtooth function
defined in terms of the (non-periodic)
stairstep function $\epsilon_{2\pi}(z)$
\footnote{
$\epsilon_{2\pi}(z)=2n+1$ for $2\pi n< z <(2n+2)\pi$
which coincides with $\hbox{sign}(z)$ for
$-2\pi< z<2\pi$.}.
$\delta_{2\pi}(z)$ denotes the periodic delta function.
Observe that the zero mode
$p_2$ enters into the algebra. 
Using (\ref{sugawara}) and the parafermion algebra
it is easy to see that the parafermions have conformal weight
one
\begin{equation}
\{T(z),W_\pm(z')\}=
-\partial_{z'}W_\pm(z')\delta_{2\pi}(z-z')+
W_\pm(z')\partial_z \delta_{2\pi}(z-z')\mp\frac{i\gamma p_2}{2\pi}
W_\pm(z')\delta_{2\pi}(z-z').
\end{equation}
One can also derive the Virasoro algebra
\begin{equation}
\{T(z),T(z')\}=-\partial_{z'}T(z')\delta_{2\pi}(z-z')
+2T(z)\partial_z\delta_{2\pi}(z-z').
\end{equation}

\section{ Quantum parafermions}

Now we turn to the quantisation of the model.
One could simply
go ahead and try to quantise the CT (\ref{ct}).
In Liouville theory locality and conformal invariance (i.e.
the requirement that the exponentials are primary)
were sufficient to uniquely fix the form of the quantum
transformation.
However, we now have double the number of degrees of freedom as
in the Liouville case and so locality and conformal invariance
are too weak.
To fix the quantum transformation we will demand that the physical
fields `close' with respect to the \it parafermions \rm
as well as the energy momentum tensor. This will be explained
in more detail in the next section.
It is clear that we need to develop a quantum version of the parafermion
algebra given in the last section.

Our starting point for the quantisation of the parafermions will be the
free field representation (\ref{periodicparas}).
The free fields can be quantised in the usual manner by defining
commutators to be $i\hbar$ times the corresponding
PB's.
Let us define parafermion operators corresponding to (\ref{periodicparas})
as
\begin{equation}
W_\pm(z)=\frac{1}{\gamma}:
\left(\eta\partial_z \phi_1(z)
\pm i\partial_z\phi_2(z)\right)e^{\pm2i\gamma\varphi_2(z)}:.
\end{equation}
With a little bit of hindsight we have included a deformation parameter
$\eta$ ($\eta\rightarrow 1$ as $\hbar\rightarrow 0$).
As usual the colons denote normal ordering.
To effect this we decompose the chiral free fields as follows
\begin{equation}\label{decomposition}
\phi_i(z)=\frac{1}{2}q_i+
\frac{p_i}{4\pi}z+
\phi_i^+(z)+\phi_i^-(z),
\end{equation}
where
\begin{equation}
\phi_i^\pm(z)=\pm\frac{i}{\sqrt{4\pi}}
\sum_{n>0}
\frac{a_{\pm n}^{(i)}}{n}e^{\mp inz}.
\end{equation}
Normal ordering is defined by moving the $\phi^+$'s to the right of
the $\phi^-$'s in all expressions.
For the zero modes Hermitian normal ordering will be understood:
\begin{equation}
:e^{2q}f(p):=e^qf(p)e^q.
\end{equation}

 It now seems that all we have to do to determine the quantum parafermion
algebra is to compute the commutators corresponding to the
Poisson brackets (\ref{PB1},\ref{PB2},\ref{PB3}).
Unfortunately, such objects appear to be ill-defined.
For example, on computing $[W_+(z),W_-(z')]$
one obtains  meaningless contributions such as
$f(z-z')\delta_{2\pi}'(z-z')$ where the derivative of $f(z)$
diverges at $z=0$.
It is conceivable that some kind of
UV renormalisation in addition to  the normal
ordering can `cure' this problem.
In \cite{fjw} a different tack was taken.
By deforming the commutator a well defined parafermion
algebra was derived.
The starting point was the following operator identity
\begin{equation}\label{opidentity}
\frac{W_+(z)W_+(z')}{e^{i\hbar \gamma^2 h^+(z-z')}}
-\frac{W_+(z')W_+(z)}{e^{-i\hbar \gamma^2 h^-(z-z')}}
=\frac{i\hbar}{2\gamma^2}\left(
\eta^2-1+\frac{\gamma^2\hbar}{\pi}\right)
:\! e^{2i\gamma\varphi(z)}
e^{2i\gamma\varphi(z')} \!:\!
\partial_z\delta_{2\pi}(z-z'),
\end{equation}
where
\begin{equation}\label{splitting}
h^\pm(z)=\frac{1}{2}h(z)\mp\log\left(4\sin^2\frac{z}{2}\right)
\end{equation}
can be viewed as the positive and negative frequency parts
of the saw-tooth function $h(z)$, respectively (see the technical
remarks at the end of this section).
The left hand side of (\ref{opidentity})
 reduces to a commutator in the limit
$\hbar\rightarrow 0$, while the right hand side is well defined.
It is however not quite what
we want, since
the operator $:e^{2i\gamma\varphi_2(z)}e^{2i\gamma\varphi_2(z')}:$
 cannot be rewritten locally in terms of the parafermions
as we would expect for a closed parafermion algebra.
We can remove the offending term altogether by imposing the 
restriction
\begin{equation}\label{qcondition}
\eta^2-1+\frac{\gamma^2\hbar}{\pi}=0,
\end{equation}
so that
\begin{equation}\label{qal1}
\frac{W_+(z)W_+(z')}{e^{i\hbar \gamma^2 h^+(z-z')}}
-\frac{W_+(z')W_+(z)}{e^{-i\hbar \gamma^2 h^-(z-z')}}
=0.
\end{equation}
This is the quantum relation corresponding to the PB (\ref{PB1}).
To check this one can expand the exact formula in powers of $\hbar$
\begin{equation}
[W_+(z),W_+(z')]-i\hbar\gamma^2
W_+(z)W_+(z')h(z-z')+O(\hbar^2)=0.
\end{equation}
Here we have used $h(z)=h^+(z)+h^-(z)$ which follows immediately from
(\ref{splitting}).
The quantum relations corresponding to the other brackets
are
\begin{eqnarray}\label{qal2}
\frac{W_-(z)W_-(z')}{e^{i\hbar\gamma^2 h^+(z-z')}}
-\frac{W_-(z')W_-(z)}{e^{-i\hbar\gamma^2 h^-(z-z')}}&=&0,\\
\label{qal3}
\frac{W_+(z)W_-(z')}{e^{-i\hbar \gamma^2 h^+(z-z')}}
-\frac{W_-(z')W_+(z)}{e^{i\hbar\gamma^2 h^-(z-z')}}&=&
\frac{i\hbar}{\gamma^2}\left(
\partial_z+\frac{i\gamma p_2}{2\pi}\right)\delta_{2\pi}(z-z'),
\\
\label{qal4}
~[p_2,W_\pm(z)]&=&\pm 2\hbar\gamma W_\pm(z).
\end{eqnarray}
As in the derivation of (\ref{qal1}) it is necessary to impose
(\ref{qcondition}) to eliminate extraneous operators.

We now turn to the energy momentum tensor. Classically this is 
 just a simple product of the parafermions,
i.e. $T(z)=\gamma^2 W_+(z)W_-(z)$.
This
 does not make sense at the quantum level.
The way out is to define $T(z)$ as a PB rather than a product.
Consider
\begin{eqnarray}\label{pbem}
\{D_z W_+(z),W_-(z')\}&=&\gamma^2D_z W_+(z) W_-(z')h(z-z')
-2T(z)\delta_{2\pi}(z-z')\nonumber \\
&&+\frac{1}{\gamma^2}D_z^2\delta_{2\pi}(z-z'),
\end{eqnarray}
where
\begin{equation}
D_z=\partial_z+\frac{i\gamma p_2}{2\pi}.
\end{equation}
This is a perfectly good, albeit unwieldy, classical
definition of $T(z)$.
The point is that we know how to `quantise' such brackets.
In fact (\ref{pbem}) is essentially the derivative of (\ref{PB2})
 whose quantum
analogue is (\ref{qal3}).
A detailed calculation yields
\begin{eqnarray}
\frac{D_z W_+(z)W_-(z')}{e^{-i\hbar \gamma^2 h^+(z-z')}}
\!\!&-&\!\!
\frac{W_-(z')D_zW_+(z)}{e^{i\hbar\gamma^2 h^-(z-z')}}
=\frac{i\hbar}{\gamma^2}\left(1+\frac{\hbar\gamma^2}{2\pi}
\right)D_z^2\delta_{2\pi}(z-z') \\
&&\!\!\!\!\!\!\!\!\!\!\!\!\!\!\!\!\!
\!\!\!\!\!\!\!\!\!\!\!\!\!\!\!\!\!\!\!\!\!\!-2i\hbar\eta^2\left(
:(\partial_z\phi_1)^2(z'):+
:(\partial_z\phi_2)^2(z'):+
\frac{\hbar\gamma}{2\pi\eta}\partial_z^2\phi_1(z')
+\frac{\hbar\gamma^2}{(4\pi\eta)^2}\right)
\delta_{2\pi}(z-z').\nonumber
\end{eqnarray}
The second entry on the right hand side corresponds to the term
$-2T(z')\delta_{2\pi}(z-z')$ of the classical PB (\ref{pbem})
suggesting the following identification
\begin{equation}
T(z)=
:(\partial_z\phi_1)^2(z):+
:(\partial_z\phi_2)^2(z):+
\frac{\hbar\gamma}{2\pi\eta}\partial_z^2\phi_1(z)
+\frac{\hbar\gamma^2}{(4\pi\eta)^2}.
\end{equation}
This is indeed the energy-momentum tensor of a free field theory with
 an additional improvement term.
One can check that this obeys a Virasoro algebra.

Classically the parafermions are primary fields of weight one.
Quantum mechanically the commutator
\begin{eqnarray}
[T(z),W_+(z')]&=&
i\hbar\left(1+\frac{\hbar \gamma^2}{2\pi}\right)
W_+(z')\partial_z \delta_{2\pi}(z-z')-
i\hbar\partial_{z'}W_+(z')\delta_{2\pi}(z-z')\nonumber \\
&&+\frac{\hbar\gamma}{2\pi}
:p_2 W_+(z'):\delta_{2\pi}(z-z')
\end{eqnarray}
shows that the quantum parafermions have the 
conformal weight $1+\hbar\gamma^2/(2\pi)$.

We end this section with some technical details
 relating to the derivation
of the quoted
operator identities  (see also the appendix
of \cite{fjw}).
The key formulae are the commutators
for the $\phi^\pm_i(z)$ entering into the decomposition (\ref{decomposition}).
A straightforward calculation gives
\begin{equation}
[\phi_i^\pm(z),\phi_j^\pm(z')]=0,~~~~
[\phi_i^\pm(z),\phi_j^\mp(z')]=-\frac{i}{4}\hbar\delta_{ij}h^\pm(z-z'),
\end{equation}
where
\begin{equation}\label{hpmdef2}
h^\pm(z)=\epsilon^\pm(z)-\frac{z}{2}.
\end{equation}
Here the $\epsilon^\pm(z)$ denote the positive and negative frequency parts
of the stairstep function, and have the Fourier representation
\begin{equation}
\epsilon^+(z)=\frac{z}{2\pi}+\frac{i}{\pi}\sum_{n>0}
\frac{e^{-in(z-i\varepsilon)}}{n},
~~~~
\epsilon^-(z)=\frac{z}{2\pi}+\frac{i}{\pi}
\sum_{n<0}\frac{e^{-in(z+i\varepsilon)}}{n}.
\end{equation}
Note that we have included a convergence factor, $\varepsilon>0$,
\it which should be retained until the end of all calculations.
\rm
In the limit $\varepsilon\rightarrow 0$ (\ref{hpmdef2}) agrees with our
earlier definition (\ref{splitting}).
We will also employ the `split' delta functions \cite{for}
$\delta^+(z)=1/2 \partial_z\epsilon^+(z)$
\begin{equation}
\delta^+(z)=\frac{1}{4\pi}+
\frac{1}{2\pi}\sum_{n>0}e^{-in(z-i\varepsilon)}
=-\frac{1}{4\pi}+\frac{1}{2\pi}\frac{1}{1-e^{-i(z-i\varepsilon)}},
\end{equation}
and similarly $\delta^-(z)=1/2 \partial_z\epsilon^-(z)$,
which have the property that $\delta^+(z)+\delta^-(z)
\rightarrow\delta_{2\pi}(z)$ as $\varepsilon\rightarrow 0$.

Our parafermion operator $W_+(z)$ can be written
\begin{equation}
W_+(z)=e_-(z)\nu(z) e_+(z),
\end{equation}
 where
\begin{equation}
e_\pm(z)=e^{i\gamma q_2} e^{2i\gamma \phi_2^\pm(z)},~~~
\nu(z)=\frac{1}{\gamma}
\left(\eta\partial_z \phi_1(z)+i\partial_z \phi_2(z)\right).
\end{equation}
A simple calculation (using 
$e^A e^B= e^B e^A e^{[A,B]}$  for $[A,B]$  complex ) gives
\begin{equation}\label{notnormalordered}
e^{-i\hbar\gamma^2 h^+(z-z')}
W_+(z)W_+(z')=e_-(z)\nu(z)e_-(z')e_+(z)\nu(z')e_+(z').
\end{equation}
The right hand side is still not normal ordered; a little algebra
yields
\begin{eqnarray}
e^{-i\hbar\gamma^2 h^+(z-z')}W_+(z)W_+(z')&=&
:W_+(z)W_+(z'): \nonumber\\
&&+e_-(z)e_-(z')[\nu^+(z),\nu^-(z')]e_+(z)e_+(z')\nonumber \\
&& +e_-(z)[\nu(z),e_-(z')]\nu(z')e_+(z)e_+(z') \nonumber \\
&&+ e_-(z)e_-(z')\nu(z)[e_+(z),\nu(z')]e_+(z') \nonumber \\
&&+ e_-(z)[\nu(z),e_-(z')][e_+(z),\nu(z')]e_+(z'),
\end{eqnarray}
where $\gamma \nu^\pm(z)
=\eta\partial_z \phi_1^\pm(z)+i\partial_z \phi_2^\pm(z)$.
Evaluating the commutators on the right hand side 
\begin{eqnarray}\label{almostfinal}
\frac{
W_+(z)W_+(z')}{e^{i\hbar\gamma^2 h^+(z-z')}}\!\!\!&=&\!\!\!
:W_+(z)W_+(z'):\!\!
+i\hbar:
\!\left(e^{2i\gamma\varphi_2(z)}W_+(z')
\!-\!W_+(z)e^{2i\gamma\varphi_2(z')}\right)\!:\!
\delta^+\!(z-z') \nonumber \\
&&\!\!\!\!\!\!\!\!\!\!\!\!\!\!\!\!\!\!\!\!+:
e^{2i\gamma\varphi_2(z)}e^{2i\gamma\varphi_2(z')}:
\left(
\frac{i\hbar}{2\gamma^2}(\eta^2-1)\partial_z
\delta^+(z-z')+\hbar^2\left(
\delta^+(z-z')\right)^2
\right).
\end{eqnarray}
Using the following identity  
\begin{equation}
\left[\delta^+(z)\right]^2=\frac{1}{(4\pi)^2}+\frac{i}{2\pi}
\partial_z\delta^+(z),
\end{equation}
the right hand side of (\ref{almostfinal}) can be written linearly
in $\delta^+(z-z')$ and its derivative.
This distribution becomes $\delta^-(z-z')$ on exchanging $z$ and $z'$.
Thus, if we take
 (\ref{almostfinal}) and subtract the equation obtained by exchanging
$z$ and $z'$, we get
\begin{eqnarray}
\frac{W_+(z)W_+(z')}{e^{i\hbar\gamma^2 h^+(z-z')}}-
\frac{W_+(z')W_+(z)}{e^{-i\hbar \gamma^2 h^-(z-z')}}\!\!\!\!&=&\!\!\!\!
i\hbar \!:\!\left(e^{2i\gamma\varphi_2(z)}W_+(z')
-W_+(z)e^{2i\gamma\varphi_2(z')}\right)\!:\!
\delta_{2\pi}(z-z')\nonumber\\
&&
\!\!\!\!\!\!\!\!\!\!\!\!\!\!\!\!\!
\!\!\!\!\!\!\!\!\!\!\!\!+\frac{i\hbar}{2\gamma^2}\left(
\eta^2-1+\frac{\hbar\gamma^2}{\pi}\right):
e^{2i\gamma\varphi_2(z)}e^{2i\gamma\varphi_2(z')}:\partial_z
\delta_{2\pi}(z-z').
\end{eqnarray}
The first term on the right hand side is zero since the prefactor
of the delta function tends to zero as $z\rightarrow z'$,
and so (\ref{opidentity}) follows immediately.
The other operator identities can be obtained in a similar manner.

\section{Physical fields and metric}

We now sketch the quantisation of the physical fields.
Classically $u$ has conformal weight zero
\begin{equation}\label{physicalweight}
\{T(z),u(z',\bar z')\}=-\partial_{z'}u(z',\bar z')\delta_{2\pi}(z-z').
\end{equation}
One can also derive the following relation
\begin{equation}\label{physicalparabracket}
\{V_+(z),u(z',\bar z')\}=\frac{i\gamma^2}{2}V_+(z)u(z',\bar z')\epsilon_{2\pi}
(z-z').
\end{equation}
We propose to use (\ref{physicalweight}) and (\ref{physicalparabracket})
as a basis for the quantisation of $u$.
The $\{V_-,u\}$ bracket is more complicated
(this does not reflect any disparity between
$V_+$ and $V_-$ since
$\{V_-,\bar u\}$
is likewise simpler 
than $\{V_+,\bar u\}$).
Quantum mechanically we expect $u$ to have a non-zero conformal weight and
that the quantum version of (\ref{physicalparabracket})
will involve the kind of deformed commutator considered in the previous 
section.

Finally, let us turn to the equal-time algebra of the physical fields.
Classically we have the `locality' relations
\begin{equation}
\{u(\sigma,\tau),u(\sigma',\tau)\}=
\{u(\sigma,\tau),\bar u(\sigma',\tau)\}=
\{\bar u(\sigma,\tau),\bar u(\sigma',\tau)\}=0,
\end{equation}
whose quantum extensions are obvious.
More interesting are the brackets
\begin{equation}
\{u(\sigma,\tau),
\dot{\bar u}(\sigma',\tau)\}=
\{\bar u(\sigma,\tau),\dot{u}(\sigma',\tau)\}
=2\gamma^2(1+u\bar u)\delta_{2\pi}(\sigma-\sigma').
\end{equation}
The coefficient of the delta is clearly related to the target space metric.
Writing $ds^2=g_{\mu\nu}dx^\mu dx^\nu$,
 the coefficient can be identified with
$g^{u \bar u}=
g^{\bar u u}$.
It would be interesting to compare the `quantum metric'
derived from the commutator
$[u(\sigma,\tau),
\dot{\bar u}(\sigma',\tau)]$
with the  quantum deformations of the target space metric
discussed in the literature \cite{dvv,ts}.

\section*{Acknowledgements}

I am grateful to G. Jorjadze and G. Weigt for their vital input
into  the work reported here.


\begin{thebibliography}{99}


\bibitem{witten} E. Witten, Phys. Rev. {\bf D44} 314 (1991)


\bibitem{gs} J.-L. Gervais and M.V. Saveliev,
Phys. Lett. {\bf B286}  271 (1992).
\bibitem{bilal} A. Bilal,
Nucl. Phys. {\bf B422} 258 (1994).

%\bibitem{fwbfo}
%P. Forgacs, A. Wipf, J. Balog, L. Feher, L. O'Raifeartaigh,
%Phys. Lett. {\bf B227} 214 (1989).

\bibitem{BCT} E. Braaten, T. Curtright, and C. Thorn,
Ann. Phys. {\bf 147} 365 (1983)

\bibitem{OW} H.J. Otto and G. Weigt,
Z. Phys.{\bf C31} 219 (1986).

\bibitem{MW} U. M\"uller and G. Weigt,
Comm. Math. Phys. {\bf 205}  421 (1999).

\bibitem{fjw} C. Ford, J. Jorjadze and G. Weigt,
\it Integration of the $SL(2,R)/U(1)$ Gauged WZNW Theory
by Reduction and Quantum Parafermions, 
\tt hep-th/0003246.\rm

\bibitem{for} C. Ford and L. O'Raifeartaigh,
Nucl. Phys. {\bf B460} 203 (1996).

\bibitem{dvv} R. Dijkgraaf, E. Verlinde, H. Verlinde,
Nucl. Phys. {\bf B371} 269 (1992).

\bibitem{ts} A.A. Tseytlin, Nucl. Phys. {\bf B399} 601 (1993).


\end{thebibliography}
\end{document}